\documentclass[]{aa}
\usepackage[varg]{txfonts}
\usepackage{graphicx}
\usepackage{color}
\begin{document}

\title{Constraint on ion--neutral drift velocity in the Class 0 protostar \object{B335} from ALMA observations}
\author{Hsi-Wei Yen\inst{\ref{inst1}}, Bo Zhao\inst{\ref{inst2}}, Patrick M. Koch\inst{\ref{inst3}}, Ruben Krasnopolsky\inst{\ref{inst3}}, Zhi-Yun Li\inst{\ref{inst4}}, Nagayoshi Ohashi\inst{\ref{inst5}}, Shigehisa Takakuwa\inst{\ref{inst6},\ref{inst3}}}

\institute{European Southern Observatory (ESO), Karl-Schwarzschild-Str. 2, D-85748 Garching, Germany \email{hyen@eso.org}\label{inst1} 
\and
Max-Planck-Institut f\"ur extraterrestrische Physik (MPE), Garching, Germany, 85748\label{inst2} 
\and
Academia Sinica Institute of Astronomy and Astrophysics, P.O. Box 23-141, Taipei 10617, Taiwan\label{inst3}
\and 
University of Virginia, Astronomy Department, Charlottesville, USA, 22904\label{inst4}
\and 
Subaru Telescope, National Astronomical Observatory of Japan, 650 North A'ohoku Place, Hilo, HI 96720, USA\label{inst5}
\and
Department of Physics and Astronomy, Graduate School of Science and Engineering, Kagoshima University, 1-21-35 Korimoto, Kagoshima, Kagoshima 890-0065, Japan\label{inst6}
}
\date{Received date / Accepted date}

\abstract
{}
{Ambipolar diffusion can cause a velocity drift between ions and neutrals. This is one of the non-ideal magnetohydrodynamcis
(MHD) effects proposed to enable the formation of large-scale Keplerian disks with sizes of tens of au. To observationally study ambipolar diffusion in collapsing protostellar envelopes, we compare here gas kinematics traced by ionized and neutral molecular lines and discuss the implication on ambipolar diffusion.} 
{We analyzed the data of the H$^{13}$CO$^+$ (3--2) and C$^{18}$O (2--1) emission in the Class 0 protostar B335 obtained with our ALMA observations. We constructed kinematical models to fit the velocity structures observed in the H$^{13}$CO$^+$ and C$^{18}$O emission and to measure the infalling velocities of the ionized and neutral gas on a 100~au scale in B335.}
{A central compact ($\sim$1$\arcsec$--2$\arcsec$) component that is elongated perpendicular to the outflow direction and exhibits a clear velocity gradient along the outflow direction is observed in both lines and most likely traces the infalling flattened envelope. With our kinematical models, the infalling velocities in the H$^{13}$CO$^+$ and C$^{18}$O emission are both measured to be 0.85$\pm$0.2~km~s$^{-1}$ at a radius of 100~au, suggesting that the velocity drift between the ionized and neutral gas is at most 0.3~km~s$^{-1}$ at a radius of 100~au in B335.}
{The Hall parameter for H$^{13}$CO$^+$ is estimated to be $\gg$1 on a 100~au scale in B335, so that H$^{13}$CO$^+$ is expected to be attached to the magnetic field. Our non-detection or upper limit of the velocity drift between the ionized and neutral gas could suggest that the magnetic field remains rather well coupled to the bulk neutral material on a 100~au scale in this source, and that any significant field-matter decoupling, if present, likely occurs only on a smaller scale, leading to an accumulation of magnetic flux and thus efficient magnetic braking in the inner envelope. This result is consistent with the expectation from the MHD simulations with a typical ambipolar diffusivity and those without ambipolar diffusion. On the other hand, the high ambipolar drift velocity of 0.5--1.0~km~s$^{-1}$ on a 100~au scale predicted in the MHD simulations with an enhanced ambipolar diffusivity by removing small dust grains, where the minimum grain size is 0.1 $\mu$m, is not detected in our observations. However, because of our limited angular resolution, we cannot rule out a significant ambipolar drift only in the midplane of the infalling envelope.
Future observations with higher angular resolutions ($\sim$0\farcs1) are needed to examine this possibility and ambipolar diffusion on a smaller scale. }


\titlerunning{Ion--neutral drift velocity in \object{B335}}
\authorrunning{H.-W. Yen et al.}

\maketitle

\section{Introduction}
With recent interferometric observations at (sub-)millimeter wavelengths, Keplerian disks with radii of tens to hundreds of au have been detected around several embedded Class 0 and I protostars \citep{Lommen08, Takakuwa12, Tobin12, Brinch13, Murillo13, Ohashi14, Chou14, Harsono14, Lindberg14, Lee14, Lee16, Yen14, Yen17, Aso15, Chou16}.
On the other hand, there is a group of Class 0 protostars with envelope rotations on a 1000~au scale that are one order of magnitude slower than other Class 0 protostars \citep{Brinch09, Yen10, Yen15a, Maret14}, 
and they do not exhibit Keplerian disks with sizes larger than 10--20~au \citep{Maury10, Oya14, Yen17}. 
The origin of the discrepancy in the gas kinematics and the disk sizes between these young protostars and others is not clear. 
It can be due to difference in their ages, initial rotation of their parental cores, and/or effects of the magnetic field \citep{Yen15a, Yen15b, Yen17}. 

Molecular clouds are magnetized \citep{Crutcher12}.
The magnetic field can slow down the gas motion in collapsing dense cores and transfer the angular momentum of the collapsing material outward, 
and consequently suppress the formation and growth of Keplerian disks around protostars \citep{Allen03}.
Ideal magnetohydrodynamical (MHD) simulations of collapsing dense cores where the rotational axis is aligned with the magnetic field show that no Keplerian disk with a size larger than 10~au can form as a result of efficient magnetic braking \citep[e.g.,][]{Galli06, Mellon08, Joos12}. 
This is consistent with the observations of some Class 0 protostars exhibiting slow envelope rotation and Keplerian disks smaller than 10--20~au, 
but it contradicts other observations showing an increasing number of Keplerian disks with radii larger than tens of au. 
In addition, 100 au scale Keplerian disks are often observed around T Tauri and Herbig Ae/Be stars \citep{Williams11}. 
Simulations have demonstrated that when non-ideal MHD effects, dissipation of   protostellar  envelopes,  initially misaligned  rotational  axis and  magnetic  field, more realistic treatment of ionization  degrees, or turbulence are considered, the efficiency of magnetic braking can be reduced to enable the formation of Keplerian disks larger than 10 au \citep{Hennebelle09, Krasnopolsky11, Li11, Li13, Machida11, Machida+11, Machida14, Dapp12, Joos12, Joos13, Santos12, Seifried12, Seifried13, Padovani13, Padovani14, Tomida15, Tsukamoto15a, Tsukamoto15b, Hennebelle16, Zhao16, Zhao18}.
Nevertheless, these mechanisms have not yet been observationally confirmed or constrained. 

Ambipolar diffusion is one of the non-ideal MHD effects proposed to enable the formation of large-scale Keplerian disks. 
With ambipolar diffusion, ions and neutrals can have a relative drift, 
and the magnetic field, which is tied to the ions, is not dragged toward the center during the collapse as quickly as in the ideal MHD case \citep{Li14}.
Thus, the magnetic flux is redistributed and is partially left in an outer region, 
and the mass-to-flux ratio increases toward the center (e.g., Fig.~7 in \citet{Dapp12} and Fig.~6 in \citet{Tomida15}). 
As a result, the efficiency of magnetic braking can be reduced to form 10~au or larger Keplerian disks depending on the magnetic diffusivity of ambipolar diffusion \citep[e.g.,][]{Dapp12, Tomida15, Tsukamoto15a, Masson16, Zhao16, Zhao18}. 
Thus, observational studies comparing the motions of ionized and neutral gas are essential to understand ambipolar diffusion in protostellar sources and its effects on the formation and growth of Keplerian disks \citep[e.g.,][]{Caselli02a,Caselli02b}. 

The isolated Bok globule B335 has an embedded Class 0 protostar (IRAS 19347+0727) at a distance of $\sim$100 pc \citep{Kee80, Kee83, Stutz08, Olofsson09}.
It is associated with a large-scale CO outflow \citep[e.g.,][]{Hirano88, Yen10, Hull14} as well as Herbig--Haro objects \citep{Reipurth92, Galfalk07}.
The signatures of the infalling motion in B335 have been observed with single-dish telescopes and interferometers on scales from hundreds to thousands of au \citep{Zho93, Zho95, Choi95, Evans05, Evans15, Saito99, Yen10, Kurono13}. 
B335 is slowly rotating with a decreasing specific angular momentum on scales from 0.1~pc to 1000~au \citep{Saito99, Yen11, Kurono13}, 
and the radial profile of its specific angular momentum flattens at an inner radius of 1000~au \citep{Yen15b}. 
No Keplerian disk with a size larger than 10~au is detected in B335 with the Atacama Large Millimeter/submillimeter Array (ALMA) observations \citep{Yen15b}. 
Near-infrared polarimetric observations show that the magnetic field on a 0.2~pc scale in B335 is tilted from the outflow axis by 35$\degr$--60$\degr$ \citep{Bertrang14}.
The structures of the magnetic field become more disordered and are misaligned with the outflow axis on a scale of a few thousand au, as revealed by single-dish and interferometric polarimetric observations at (sub-)millimeter wavelengths \citep{Wolf03, Davidson11, Chapman13, Hull14}.
Because of its slowly rotating inner envelope and absence of a large disk, B335 is a promising candidate with efficient magnetic braking \citep{Yen15b}.
Therefore, B335 is an excellent target for probing the absence or existence of ambipolar diffusion and for studying the effects of the magnetic field in the process of star formation.

We here report our observational results of the H$^{13}$CO$^+$ (3--2; 260.255342 GHz) emission in B335 obtained with ALMA, 
and we compare the gas kinematics traced by the H$^{13}$CO$^+$ (3--2) line with the C$^{18}$O (2--1) results presented in \citet{Yen15b}.  
C$^{18}$O and H$^{13}$CO$^+$ are chemically similar \citep{Lee04, Aikawa08}, 
and both are expected to be abundant in the inner envelope on a scale of hundreds of au in B335 \citep{Evans05}.
The C$^{18}$O (2--1) line has an upper energy level of 16~K and a critical density of $\sim$10$^{4}$~cm$^{-3}$, 
and the H$^{13}$CO$^+$ (3--2) line has an upper energy level of 25~K and a critical density of $\sim$10$^{5}$~cm$^{-3}$. 
The upper energy levels and critical densities of both lines are below the temperature and density in the protostellar envelope on a scale within a few hundred au in B335 \citep{Harvey03, Shirley11}, 
and both lines are expected to be optically thin on this scale based on the typical column densities of C$^{18}$O and H$^{13}$CO$^+$ in protostellar envelopes on a scale of hundreds of au \citep{Hogerheijde98}. 
Therefore we here adopt these two lines to trace the kinematics of the ionized and neutral gas in the protostellar envelope in B335, 
and we compare the measured infalling velocities from these two lines and discuss the implication of our results on the effects of ambipolar diffusion in B335.

\section{Observations}\label{ob}
The data of B335 presented in this paper were obtained with the ALMA observations with 40 antennas during the cycle-3 observing period on May 23 and June 3, 2016.
The array configuration was C36-4 with the shortest baseline length of $\sim$15 m ($\sim$13 k$\lambda$).
Because of different array configurations, the shortest baseline length in the H$^{13}$CO$^+$ observations is half of that in the ALMA C$^{18}$O observations with the C34-6 configuration \citep{Yen15b}. 
With this shortest baseline length, our H$^{13}$CO$^+$ observations have a largest recoverable angular scale of 7$\arcsec$ at a 50\% level (Wilner \& Welch 1994).
The pointing center was $\alpha$(J2000) = $19^{h}37^{m}00\fs89$, $\delta$(J2000) = $7\degr34\arcmin9\farcs6$. 
The on-source integration time on B335 was $\sim$80 minutes. 
The correlator was configured in the frequency division mode, 
and a spectral window with a bandwidth of 117.2 MHz was assigned to the H$^{13}$CO$^+$ emission with 960 channels, resulting in a channel width of 122.1 kHz. 
The 1.2 mm continuum was observed simultaneously with a total bandwidth of 2 GHz. 
We have confirmed that the peak position of the 1.2 mm continuum emission is consistent with the 1.3 mm continuum emission in \cite{Yen15b}, 
so that there is no relative positional offset between the two data sets. 
The position of the continuum peak, which is the same as the pointing center, was adopted as the protostellar position.
J1935+2031 (0.75 Jy at 260.2 GHz) and J2035+1056 (0.31 Jy at 260.2 GHz) were observed as gain calibrators in the first and second observations, respectively.  
J2148+0657 was observed as a bandpass and flux calibrator. 
Calibration of the raw visibility data was performed with the standard reduction script for the cycle-3 data, which uses tasks in Common Astronomy Software Applications \citep[CASA;][]{McMullin07} of version 4.5.3.
The image of the H$^{13}$CO$^+$ emission was generated with the Briggs robust parameter of 0 from the calibrated visibility data and CLEANed with the CASA task ``clean'' at a velocity resolution of 0.17~km~s$^{-1}$.
This velocity resolution is the same as that in the C$^{18}$O image obtained from \citet{Yen15b}. 
The achieved synthesized beam is 0\farcs5 $\times$ 0\farcs4 with a position angle (PA) of 111\degr, 
which is a factor of 1.5 larger than that of the C$^{18}$O image. 
The achieved rms noise is 3.3~mJy~beam$^{-1}$ per channel, comparable to that of the C$^{18}$O observations (3.8~mJy~beam$^{-1}$ per channel).

\section{Results}

\begin{figure*}
\centering
\includegraphics[width=18cm]{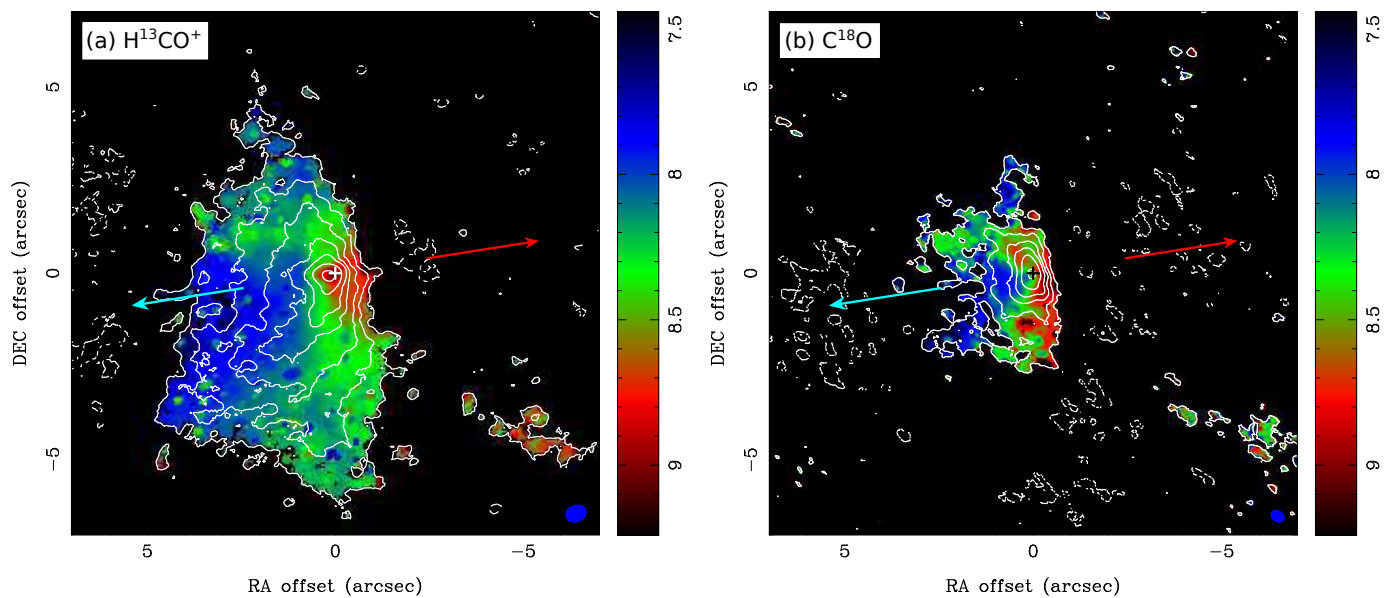}
\caption{Moment 0 (contours) overlaid on moment 1 (color) maps of the (a) H$^{13}$CO$^+$ (3--2) and (b) C$^{18}$O (2--1) emission in B335 obtained with the ALMA observations. Color scales are in units of km s$^{-1}$ in the LSR frame. The color scales are the same in (a) and (b). Crosses denote the protostellar position. Red and blue arrows show the direction of the blue- and redshifted outflows, respectively. Contour levels in (a) are 3$\sigma$, 6$\sigma$, 10$\sigma$, 15$\sigma$, 20$\sigma$, 30$\sigma$, and 45$\sigma,$ and in (b) are 3$\sigma$, 6$\sigma$, 10$\sigma$, 15$\sigma$, 25$\sigma$, and 40$\sigma$, where 1$\sigma$ is 3.3 and 3.4~mJy~beam$^{-1}$~km~s$^{-1}$, respectively.}\label{moment}
\end{figure*}

Figure \ref{moment} presents the total integrated-intensity (moment 0) and mean intensity-weighted velocity (moment 1) maps of the H$^{13}$CO$^+$ and C$^{18}$O emission. 
The maps of the C$^{18}$O emission have been presented in \citet{Yen15b} and are also shown here for a direct comparison with the H$^{13}$CO$^+$ maps. 
The distributions of both emission lines are centrally peaked at the protostellar position, 
and the H$^{13}$CO$^+$ emission exhibits a blueshifted extension on a scale of 5$\arcsec$ (500 au) toward the east. 
A similar blueshifted extension toward the east is also observed in the C$^{18}$O emission, but on a smaller scale. 
The observed H$^{13}$CO$^+$ emission is more extended than the C$^{18}$O emission because the shortest baseline length in the H$^{13}$CO$^+$ observations is almost half of that in the C$^{18}$O observations. 
The central parts of both emission lines are more elongated along the north--south direction. 
By fitting a two-dimensional Gaussian distribution to the central part of these emission lines, where the integrated intensity is $>$20$\sigma$ in the H$^{13}$CO$^+$ emission and $>$10$\sigma$ in the C$^{18}$O emission, 
the PA of the elongation of the H$^{13}$CO$^+$ and C$^{18}$O emission around the protostar is measured to be $-12$\degr$\pm$19\degr and 17\degr$\pm$11\degr, respectively.
These elongations are perpendicular to the outflow direction with a PA of 99$\degr$ \citep{Hull14} and are more aligned with the flattened envelope with a PA of 16$\degr$ on a 100 au scale traced by the 1.3 mm continuum \citep{Yen15b}.
The H$^{13}$CO$^+$ and C$^{18}$O emission lines both exhibit a clear velocity gradient along the east--west direction, where the eastern side is more blueshifted and the western side is more redshifted. 
The direction of this velocity gradient is consistent with the associated outflow. 
As discussed in \citet{Yen15b},
the blueshifted extended emission toward the east is most likely related to the blueshifted outflow, 
and the velocity gradient in the central part, where the emission is elongated along the direction of the flattened envelope, can be due to the infalling motion in the envelope.  

\begin{figure}
\centering
\includegraphics[width=7.5cm]{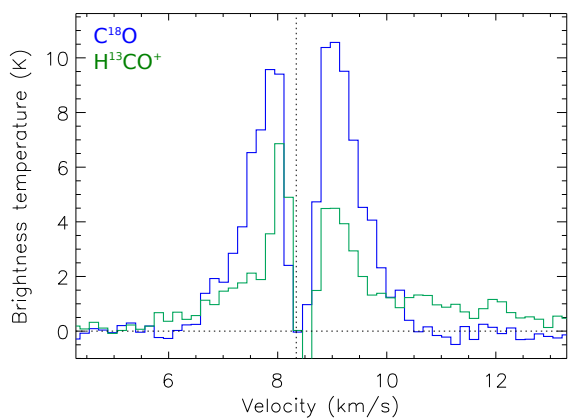}
\caption{Spectra of the C$^{18}$O and H$^{13}$CO$^+$ emission at the protostellar position obtained with our ALMA observations, shown in blue and green histograms. The data were first convolved with the same circular beam with a size of 0\farcs5 to extract the spectra. A vertical dotted line denotes the systemic velocity of 8.34~km~s$^{-1}$.}\label{spec}
\end{figure}

\begin{figure*}
\centering
\includegraphics[width=18cm]{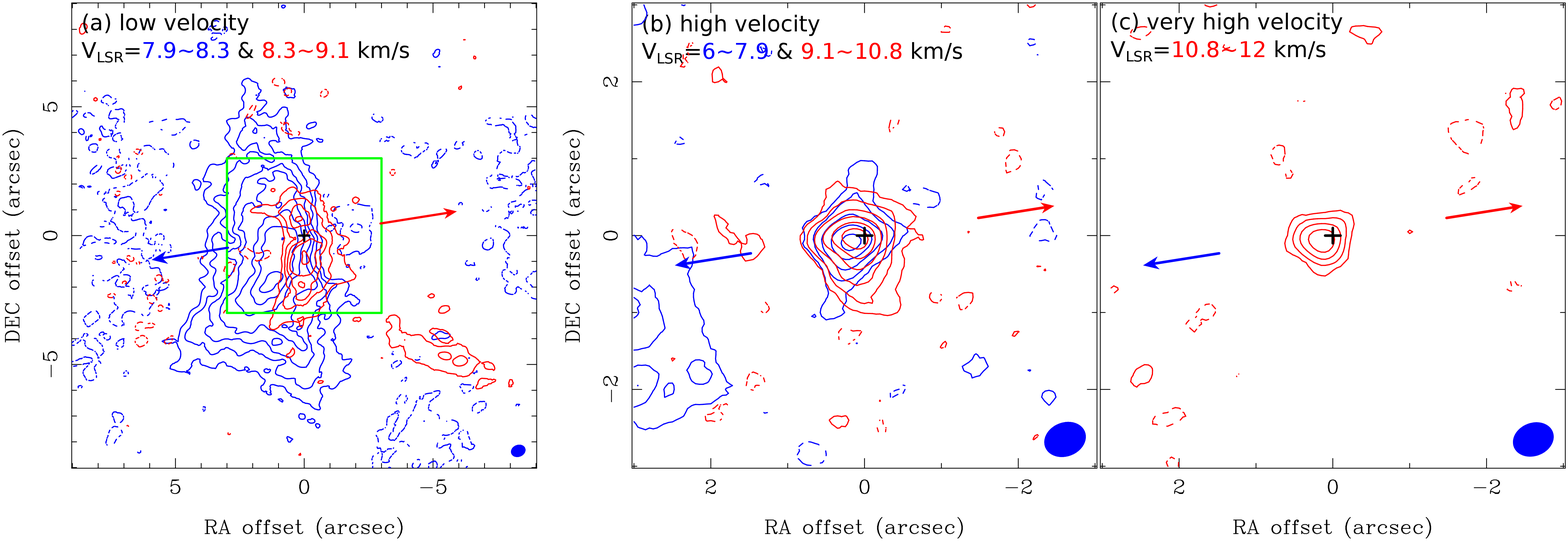}
\caption{Moment 0 maps of the H$^{13}$CO$^+$ emission integrated over different velocity regimes, (a) low velocities $V_{\rm LSR} = 7.9\mbox{--}8.3$ \& $8.3\mbox{--}9.1$~km~s$^{-1}$, (b) high velocity $V_{\rm LSR} = 6\mbox{--}7.9$ \& 9.1--10.8~km~s$^{-1}$, and (c) very high velocity $V_{\rm LSR} = 10.8\mbox{--}12$~km~s$^{-1}$. Blue and red contours present the blue- and redshifted emission, respectively. Crosses denote the protostellar position. Red and blue arrows show the direction of the blue- and redshifted outflows, respectively. A green box in (a) presents the image size of (b) and (c). Filled blue ellipses show the size of the synthesized beam of 0\farcs5 $\times$ 0\farcs4. Contour levels in (a) are 5$\sigma$, 10$\sigma$, 15$\sigma$, and 20$\sigma$ and then in steps of 10$\sigma$, in (b) are 3$\sigma$, 6$\sigma$, 10$\sigma$, 15$\sigma$, 20$\sigma$, and 25$\sigma$ and then in steps of 10$\sigma$, and in (c) are from 3$\sigma$ in steps of 3$\sigma$, where 1$\sigma$ is 1.1, 1.8, and 1.6~mJy~beam$^{-1}$~km~s$^{-1}$, respectively.}\label{h13comap}
\end{figure*}

Figure \ref{spec} presents the spectra of the C$^{18}$O and H$^{13}$CO$^+$ emission at the protostellar position. 
To compare the spectra of the two lines obtained with different observations, 
both data were first convolved with the same circular beam with a size of 0$\farcs$5, 
and the intensity is converted into brightness temperature in units of K.
Both emission lines show intensity peaks at similar velocities, $V_{\rm LSR}$ of 8 and 9~km~s$^{-1}$ as well as a dip close to the systemic velocity of $V_{\rm LSR} = 8.34$~km~s$^{-1}$.
The H$^{13}$CO$^+$ emission additionally shows a redshifted line wing at $V_{\rm LSR} > 10.5$ ~km~s$^{-1}$.
We integrated the H$^{13}$CO$^+$ emission over the three different velocity regimes, low velocity ($V_{\rm LSR} = 7.9\mbox{--}8.3$  \& $8.3\mbox{--}9.1$~km~s$^{-1}$), high velocity ($V_{\rm LSR} = 6\mbox{--}7.9$ \& 9.1--10.8~km~s$^{-1}$), and very high velocity ($V_{\rm LSR} = 10.8\mbox{--}12$~km~s$^{-1}$), shown in Fig.~\ref{h13comap}. 
Similar maps of the C$^{18}$O emission have been presented in \citet{Yen15b}.
The extended emission is primarily observed at the low velocities, 
and the emission is elongated along the north-south direction with a blueshifted extension toward the east. 
A separated redshifted  component is seen at $\sim$7$\arcsec$ southwest to the protostar, possibly associated with the wall of the outflow cavity. 
Thus, at the low velocities, the H$^{13}$CO$^+$ emission likely traces the flattened envelope with a possible contamination from the outflow.
On the other hand, at high velocities, 
blue- and redshifted compact components with a size of $\sim$2$\arcsec$ (200 au) are clearly seen and are not surrounded by any extended structures, suggesting that the emission becomes less contaminated by the outflow. 
There is an additional component at even higher velocities with a relative velocity ($\Delta V$) of 2.5--4 km s$^{-1}$ in the H$^{13}$CO$^+$ emission. 
This component is not observed in the C$^{18}$O emission, and neither does it have a blueshifted counterpart.  
Because this very high velocity component is not resolved with our observations, its origin is not clear. 
If it originates from the infalling motion in the inner envelope, 
its distance to the protostar is estimated to be $\sim$5--10 AU on the assumption of a protostellar mass of 0.04 $M_\sun$ \citep{Yen15b}.
However, in this case, a blueshifted counterpart is expected, which is not observed.  
The other possibility is that this very high velocity component in the H$^{13}$CO$^+$ emission is related to the jets in B335. 
Emission at an extremely high velocity up to $\sim$20~km s$^{-1}$ tracing the jets has been observed in the $^{12}$CO (2--1) line with the SMA \citep{Yen10}. 

\begin{figure*}
\centering
\includegraphics[width=16cm]{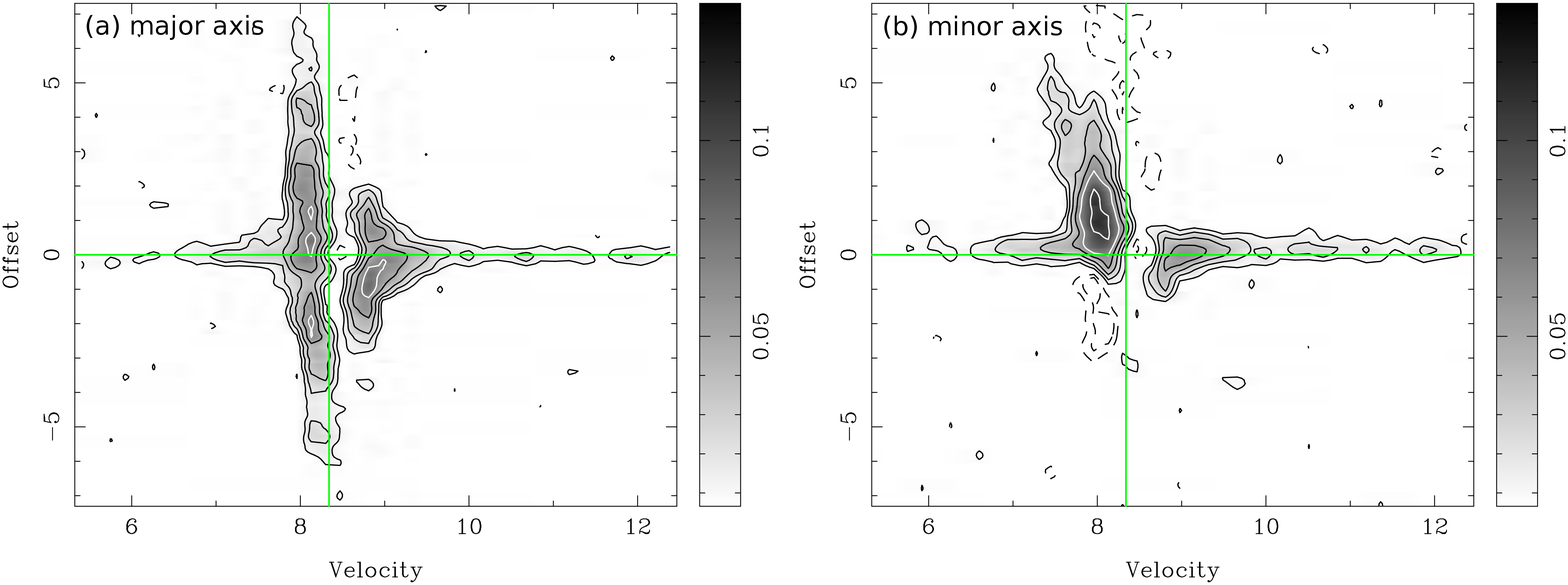}
\caption{Position--velocity diagram of the H$^{13}$CO$^+$ emission along the (a) major and (b) minor axes of the flattened envelope passing through the protostellar position. Velocity axes are in units of km~s$^{-1}$ in the LSR frame, and offset axes are in arcsecond with respect to the protostellar position. Vertical and horizontal green lines denote the systemic velocity of 8.34 km s$^{-1}$ and the protostellar position, respectively. Contour levels are 3$\sigma$, 6$\sigma$, 10$\sigma$, 15$\sigma$, and 20$\sigma$ and then in steps of 10$\sigma$, where 1$\sigma$ is 3.3~mJy~beam$^{-1}$.}\label{pv}
\end{figure*}

Figure \ref{pv} presents the position--velocity (PV) diagrams along the major and minor axes of the flattened envelope passing through the protostellar position. 
The PA of the major axis of the flattened envelope has been measured to be 16$\degr$ from the 1.3 mm continuum at an angular resolution of 0\farcs3 \citep{Yen15b}.
The PV diagrams of the H$^{13}$CO$^+$ emission are morphologically similar to those in the C$^{18}$O emission \citep[Fig.~4 in ][]{Yen15b}.
The PV diagrams show that the velocity width increases as radii decrease, 
and in the central 2$\arcsec$ region along the major axis, there are both blue- and redshifted components at the same offsets. 
These features are consistent with the signature of an edge-on infalling envelope \citep[e.g., Fig.~10 in][]{Ohashi97} and can be explained with accelerating infalling gas behind and in front of the protostar along the line of sight. 
On the other hand, there is no clear positional offset between the blue- and redshifted emission in the central H$^{13}$CO$^+$ component ($|{\rm offset}| < 1\arcsec$) in both PV diagrams, different from the C$^{18}$O observations, which show clear velocity gradients along both axes \citep{Yen15b}. 
The reason might be the lower angular resolution of the H$^{13}$CO$^+$ observations.
The linear velocity gradient along the major axis of the envelope in the C$^{18}$O emission is measured to be 40.5~km~s$^{-1}$~arcsec$^{-1}$ from the high-velocity ($|\Delta V| >1.2$~km~s$^{-1}$) channels in the C$^{18}$O PV diagram \citep{Yen15b}. 
From this linear velocity gradient, the peak offset between blueshifted and redshifted emission at $\Delta V$ of $\pm$1.2~km~s$^{-1}$ is expected to be 0\farcs06. 
At $\Delta V$ of $\pm$1.2~km~s$^{-1}$, 
the signal-to-noise ratios (S/N) of the H$^{13}$CO$^+$ emission in our observations is $\sim$10 at the peaks. 
The relative positional accuracy of our observations at these velocities is approximately the angular resolution of 0\farcs5 divided by the S/N of 10, yielding 0\farcs05, which is comparable to the expected positional offset from the velocity gradient.  
Therefore, the angular resolution of our H$^{13}$CO$^+$ observations is not sufficient to resolve the velocity gradient observed in the C$^{18}$O emission.
We are thus not able to detect the slow envelope rotation on a 100~au scale, which we did observe in the C$^{18}$O and SO emission, with our H$^{13}$CO$^+$ observations. 

\section{Kinematical models of infalling and rotational motions}\label{kmod}
The central component with a radius of 2$\arcsec$ in the H$^{13}$CO$^+$ emission is elongated along the major axis of the flattened envelope and is perpendicular to the outflow direction (Fig.~\ref{moment}). 
Its velocity features can be explained with infalling motion in an edge-on envelope (Fig.~\ref{pv}). 
Thus, the central H$^{13}$CO$^+$ emission likely traces the infalling flattened envelope around B335, 
and no envelope rotation is detected in the H$^{13}$CO$^+$ line.
To measure the infalling velocity in the H$^{13}$CO$^+$ line, 
we constructed kinematical models of an infalling flattened envelope to fit the observed velocity structures in the PV diagrams of the H$^{13}$CO$^+$ emission.  
In order to compare the infalling velocities measured in the H$^{13}$CO$^+$ and C$^{18}$O lines, 
we also reanalyzed the C$^{18}$O data with the same kinematical models adopted in this work. 

The model envelope has three dimensions and is assumed to have power-law density and temperature profiles.  
Its number density within 60$\degr$ from the polar axis is artificially set to zero to mimic the outflow cavity and the flattened envelope, meaning that the thickness of the model envelope $h(r) = r \sin 30\degr$, identical to \citet{Yen15b}. 
This radial profile of the envelope thickness was determined based on the aspect ratios of the 1.3 mm continuum emission observed with the SMA and ALMA because B335 is almost edge on.
We fixed the temperature profile in the model envelope because the temperature cannot be well constrained with only one transition in the H$^{13}$CO$^+$ and C$^{18}$O lines. 
The power-law index of the temperature profile is adopted to be $-0.4$ \citep{Shirley00, Shirley11}, 
and the temperature is adopted to be 38~K at a radius of 100 au \citep{Evans15}. 
As discussed in \citet{Yen17}, 
because the emission is optically thin, 
the fitted velocity profile is not sensitive to the adopted temperature profile, 
and adopting a different temperature profile results in a different fitted density profile, 
which compensates for the change in the line intensity due to the different temperature profile. 
Thus, the density and temperature profiles of the model envelope are described as
\begin{equation}
n(r) = n(R_0)\cdot (\frac{r}{R_0})^{p},
\end{equation}
and 
\begin{equation}
T(r) = 38 \cdot (\frac{r}{100~{\rm au}})^{-0.4}\ {\rm K},
\end{equation}
where $n$ is the number density of C$^{18}$O or H$^{13}$CO$^+$ and $R_0$ is the characteristic radius adopted to be 100 au.
The outer radius is adopted to be 7$\arcsec$ (700 au), so the diameter of the model envelope is twice larger than the maximum recoverable angular scale of the H$^{13}$CO$^+$ observations. 
The model envelope is assumed to be free-falling with a constant specific angular momentum. 
The radial profiles of the infalling and rotational velocity ($V_{\rm in}$ and $V_{\rm rot}$) are described as 
\begin{equation}\label{vin}
V_{\rm in}(r) = V_{\rm in}(R_0) \cdot (\frac{r}{R_0})^{-0.5},
\end{equation}
and 
\begin{equation}\label{vin}
V_{\rm rot}(r) = \frac{j}{r}\sin\theta,
\end{equation}
where $j$ is the specific angular momentum and $\theta$ is the angle between the radius and the polar axis. 
As described below, 
the H$^{13}$CO$^+$ and C$^{18}$O data are fitted separately.
The best-fit $n(R_0)$, $p$, $V_{\rm in}(R_0)$, and $j$ can be different for the H$^{13}$CO$^+$ and C$^{18}$O emission.
We adopt the latest estimates of the distance of 100 pc \citep{Olofsson09} and an inclination angle of 87$\degr$ \citep{Stutz08} for B335 to compute the model images. 

\begin{figure*}
\centering
\includegraphics[width=16cm]{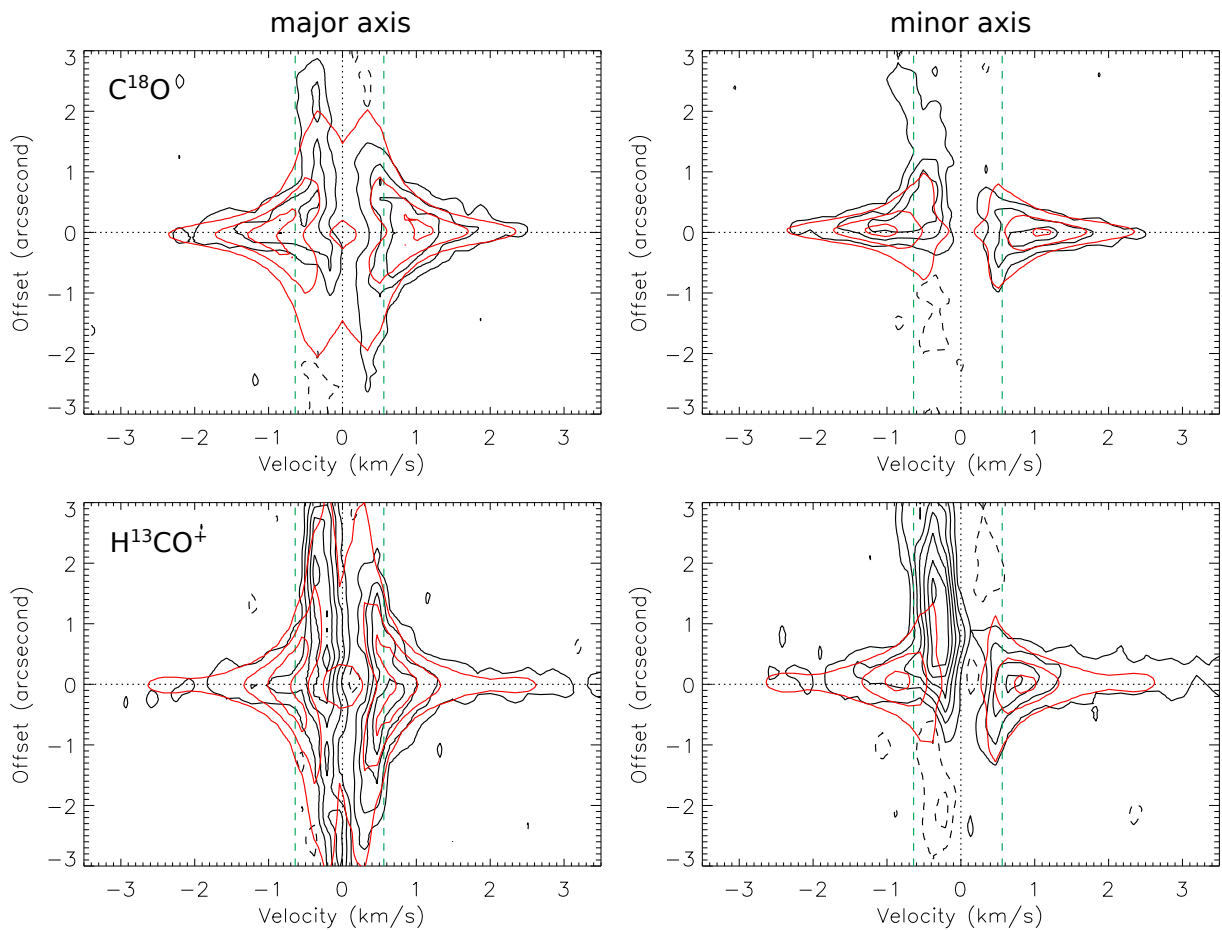}
\caption{Observed (black) and best-fit model (red; with parameters in Table \ref{fitting}) position--velocity diagrams of the C$^{18}$O (upper panels) and H$^{13}$CO$^+$ (lower panels) emission along the major (left panels) and minor (right panels) axes of the flattened envelope passing through the protostellar position. Vertical and horizontal dotted lines denote the systemic velocity of 8.34 km s$^{-1}$ and the protostellar position, respectively. Vertical green dashed lines show the velocity range that was not included in our model fitting, $V_{\rm LSR}$ of 7.7--8.9~km~s$^{-1}$, because of a possible contamination from the outflow. Contour levels are from 3$\sigma$ in steps of 5$\sigma$, where 1$\sigma$ is 3.8~mJy~beam$^{-1}$ in the C$^{18}$O emission and is 3.3~mJy~beam$^{-1}$ in the H$^{13}$CO$^+$ emission.}\label{modelpv}
\end{figure*}

We first performed the $\chi^2$ fitting on the C$^{18}$O PV diagrams along the major and minor axes. 
The model images in the C$^{18}$O emission were computed on the assumption of the local thermal equilibrium (LTE) condition because the C$^{18}$O  (2--1) line is expected to be thermalized with the typical physical condition ($n_{\rm H_2} > 10^4$~cm$^{-3}$) in protostellar envelopes on a scale of a few hundred au \citep[e.g.,][]{Shirley00}.  
Then we simulated ALMA observations of the model images with the same array configuration and coverage of hour angle as our real observations.  
Therefore, the effects of different $uv$ sampling and angular resolutions between the H$^{13}$CO$^+$ and C$^{18}$O observations are included in our analysis.
We generated synthetic images from the simulated visibility data with the same imaging parameters as \citet{Yen15b} and extracted PV diagrams from the synthetic images. 
The channels at the low velocities of $V_{\rm LSR} = 7.7\mbox{--}8.9$~km~s$^{-1}$ (vertical dashed lines in Fig.~\ref{modelpv}) are excluded in the fitting to have minimal contamination from the extended structures and the outflows. 
As demonstrated in \citet{Yen15b},  the envelope rotation is so slow that it does not affect the measurements of the infalling velocity. 
In addition, the signature of the envelope rotation is not detected in the H$^{13}$CO$^+$ emission. 
Thus, the fitting was performed with two fixed $j$, $3 \times10^{-5}$~km~s$^{-1}$~pc, as derived from \citet{Yen15b}, with the correction of the different adopted distances, and 0, meaning no rotation. 
We confirmed that the best-fit $V_{\rm in}(R_0)$ is the same with these two $j$ because the gas kinematics is dominated by the infalling motion. 

\begin{table}
\caption{Best-fit parameters of kinematical models}\label{fitting}
\centering
\begin{tabular}{lll}
\hline\hline
& C$^{18}$O (2--1) & H$^{13}$CO$^+$ (3--2) \\
\hline 
$V_{\rm in}$(100~au) & 0.85$\pm$0.2~km~s$^{-1}$ & 0.85$\pm$0.2~km~s$^{-1}$ \\
$n$(100~au) & 2.7$\pm$1.2~cm$^{-3}$ & (4$\pm$1)~$\times$~10$^{-4}$~cm$^{-3}$\\
$p$ & $-2.1$$\pm$0.5 & $-2.1$$\pm$0.5\\
\hline
\end{tabular}
\end{table}

\begin{figure}
\centering
\includegraphics[width=7.5cm]{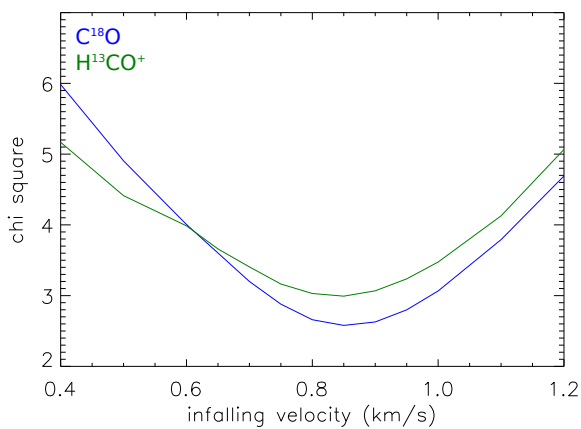}
\caption{$\chi^2$ as a function of infalling velocity at a radius of 100~au of our kinematical models for C$^{18}$O (blue curve) and H$^{13}$CO$^+$ (green curve). For each infalling velocity, $\chi^2$ presented here is the smallest $\chi^2$ achieved by varying all the other parameters. The minimum $\chi^2$ for the neutral and ionized lines are at 0.85~km~s$^{-1}$.}\label{chicurve}
\end{figure}

The best-fit results of the C$^{18}$O data are shown in Fig.~\ref{modelpv} and are listed in Table \ref{fitting}. 
The $\chi^2$ of our best-fit model is 2.6. 
The uncertainties of our best-fit parameters are estimated from the parameter ranges of the models having $\chi^2$ less than the minimum $\chi^2$ + 1 (Table \ref{fitting}).
Figure \ref{chicurve} presents $\chi^2$ as a function of $V_{\rm in}$(100~au) from our fitting results.
In Appendix \ref{sigmod}, we present PV diagrams extracted from the synthetic images of our kinematical models with different $V_{\rm in}$(100~au) to demonstrate the dependence of the velocity structures on the infalling velocity.
These comparisons show that the infalling velocity at a radius of 100~au is most likely within the range of 0.85$\pm$0.2~km~s$^{-1}$.

The best-fit C$^{18}$O number density is 2.7~cm$^{-3}$. 
On the assumption of a typical C$^{18}$O abundance of $3\times10^{-7}$ relative to H$_2$ \citep{Frerking82}, 
the H$_2$ number density is estimated to be $9\times10^6$~cm$^{-3}$. 
In addition, the C$^{18}$O abundance in the protostellar envelope in B335 is possibly lower than  the typical value in the interstellar medium (ISM) by a factor of ten \citep{Evans05,Yen10}, 
and this leads to an even higher H$_2$ number density. 
The H$^{13}$CO$^+$ (3--2) emission is also expected to be thermalized at this H$_2$ number density and the temperature of 38~K, as calculated with the non-LTE radiative transfer code {\it RADEX} \citep{radex07}.
Thus, the model images in the H$^{13}$CO$^+$ emission were also computed with the LTE condition.  
We generated synthetic PV diagrams following the same process as described above,  
and performed the $\chi^2$ fitting on the H$^{13}$CO$^+$ PV diagrams along the major and minor axes. 
The best-fit results of the H$^{13}$CO$^+$ emission are also shown in Fig.~\ref{modelpv} and are listed in Table \ref{fitting}.
The $\chi^2$ of our best-fit model for H$^{13}$CO$^+$ is 3, 
and $\chi^2$ as a function of $V_{\rm in}$(100~au) from our fitting results is shown in Fig.~\ref{chicurve}.
For comparison with the best-fit model, PV diagrams extracted from the synthetic images of the kinematical models with different $V_{\rm in}$(100~au) for H$^{13}$CO$^+$ are presented in Appendix \ref{sigmod}.

In addition, to examine the influence of the uncertainty in the PA of the major and minor axes on our fitting results,
we have changed the PA by $\pm$10$\degr$ and performed the same $\chi^2$ fitting. 
The resulting best-fit infalling velocities remain unchanged. 
Thus, our results are not sensitive to the uncertainty in the PA of the major and minor axes. 
Furthermore, we note that in the non-ideal MHD simulations \citep[e.g.,][]{Li11, Zhao18}, the radial profile of infalling velocity can be shallower than our assumed profile, $r^{-0.5}$, and the velocity profile of ions can be almost flat or even decrease with decreasing radii when the matter and the magnetic field are decoupled.
Nevertheless, our data show that the line widths of the C$^{18}$O and H$^{13}$CO$^+$ emission increase with decreasing radii (Fig.~\ref{modelpv}), suggesting that the velocity profiles of both neutral and ionized gas increase with decreasing radii on a scale of 100~au. 
To test the influence of our assumption of the velocity profile, 
we have also adopted a shallower radial profile of infalling velocity by changing the power-law index by a factor of two to be $r^{-0.25}$ and performed the same $\chi^2$ fitting. 
The difference between the best-fit $V_{\rm in}$(100~au) with the two different velocities profiles is 0.1~km~s$^{-1}$, less than our uncertainty.
Therefore, our results are also not sensitive to the assumption of the velocity profile.

The best-fit infalling velocities in the C$^{18}$O and H$^{13}$CO$^+$ emission are identical within the uncertainty. 
The difference in the infalling velocities in the C$^{18}$O and H$^{13}$CO$^+$ emission is estimated to be less than 0.3~km~s$^{-1}$ at a radius of 100~au from the error propagation of their uncertainties in $V_{\rm in}$(100~au).
This measured infalling velocity corresponds to a protostellar mass of $\sim$0.04~$M_\sun$ on the assumption that the infalling motion is free fall. 
This is consistent with \citet{Yen15b} but is lower than the estimate from fitting the inverse P-Cygni profile in the HCN and HCO$^+$ emission with the inside-out collapse model, 0.2~$M_\sun$, by \citet{Evans15}. 
In \citet{Evans15}, there are high-velocity wings in the spectra of the HCN and HCO$^+$ lines, which need to be explained with the infalling velocity caused by the higher protostellar mass. 
The lines analyzed in \citet{Evans15} are optically thicker than the lines analyzed in this work, 
and these high-velocity wings are not detected in the C$^{18}$O line. 
There is a hint of a redshifted high-velocity wing in the H$^{13}$CO$^+$ emission, but there is no blueshifted counterpart. 
Observations at higher angular resolutions to reveal the origins of these high-velocity wing and to resolve the Keplerian disk in B335 are needed to clarify the difference in the estimated protostellar mass. 
Nevertheless, our results showing that the infalling velocities are identical within the uncertainty are not affected by the discrepancy in the estimated protostellar mass because the observed velocity structures in the C$^{18}$O and H$^{13}$CO$^+$ emission can be explained with the same kinematical models.  

\section{Discussion}
By fitting our kinematical models to the observed velocity structures in the PV diagrams of the C$^{18}$O and H$^{13}$CO$^+$ emission, 
we have found that there is no detectable difference in the infalling velocities traced by the two emission lines, 
and the velocity difference between the C$^{18}$O and H$^{13}$CO$^+$ gas is estimated to be at most 0.3~km~s$^{-1}$ at a radius of 100~au in B335. 

If H$^{13}$CO$^+$ is attached to the magnetic field on a 100~au scale in B335, the velocity difference between H$^{13}$CO$^+$ and C$^{18}$O can trace the ambipolar drift velocity, which is the relative velocity between the magnetic field and the neutrals.  
Whether H$^{13}$CO$^+$ is attached to the magnetic field depends on the ratio between the Lorenz force and the drag force on H$^{13}$CO$^+,$ and it can be evaluated with the Hall parameter \citep[$\beta_{i,{\rm H_2}}$;][]{Zhao16}, 
\begin{equation}\label{betaeq}
\beta_{i,{\rm H_2}} = \frac{Z_i e B}{m_i c}\cdot \frac{m_i+m_{\rm H_2}}{\mu m_{\rm H} n({\rm H_2}) \langle \sigma_i v \rangle_{i,{\rm H_2}}}, 
\end{equation}
where $m_i$ and $Z_i e$ are the mass and the charge of H$^{13}$CO$^+$, $B$ is the magnetic field strength, $c$ is the speed of light, $\mu$ is the mean molecular weight of 2.36, $m_{\rm H}$ and $m_{\rm H_2}$ are the masses of atomic and molecular hydrogen, $n({\rm H_2})$ is the number density of H$_2$, and $\langle \sigma_i v \rangle_{i,{\rm H_2}}$ is the momentum transfer rate coefficient. 
When $\beta_{i,{\rm H_2}} \gg 1$, meaning that the Lorenz force dominates over the drag force, H$^{13}$CO$^+$ is attached to the magnetic field. 
At a radius of 100~au in B335, $n({\rm H_2})$ is estimated to be $9 \times 10^6$~cm$^{-3}$ with the ISM C$^{18}$O abundance in Section~\ref{kmod}. 
$\langle \sigma_i v \rangle_{i,{\rm H_2}}$ is a function of temperature and velocity difference between ions and neutrals ($v_{\rm d}$), 
and $\langle \sigma_i v \rangle_{i,{\rm H_2}} \propto {v_{\rm d}}^{0.6}$ for HCO$^+$ \citep{Pinto08}.
Our results suggest that $v_{\rm d}$ is less than 0.3~km~s$^{-1}$.
With the formulae in \citet{Pinto08}, 
$\langle \sigma_i v \rangle_{i,{\rm H_2}}$ is estimated to be $1.2 \times 10^9$~cm$^3$~s$^{-1}$ at a temperature of 38~K and $v_{\rm d}$ of 0.3~km~s$^{-1}$.
With the method described by \citet{Chandrasekhar53},  
$B$ has been estimated to be 10--40~$\mu$G on a 0.1 pc scale from the infrared polarimetric observations \citep{Bertrang14} and to be 134$^{+46}_{-39}$~$\mu$G on a 4000 au scale from the polarized thermal dust emission at the submillimeter wavelengths \citep{Wolf03} in B335.
On the other hand, 
if the simple power-law relation between the magnetic field strength and the density in molecular clouds, $B = 0.143 \times n({\rm H_2})^{0.5}$~$\mu$G, is valid in B335 \citep{Nakano02}, 
$B$ is estimated to be 400~$\mu$G on a 100~au scale with our estimated $n({\rm H_2})$.
In other protostellar sources, $B$ on a scale of a few hundred au has been estimated to be even higher with $\sim$5~mG from the polarized thermal dust emission \citep{Girart06, Hull17}.
It thus seems very plausible to have $B$ of a few hundred $\mu$G in B335 on a 100~au scale. 
Still, with a conservative value of $B > 100$~$\mu$G, $\beta_{i,{\rm H_2}}$ is estimated to be $>$20 on a 100~au scale.
Therefore, on the assumptions of the typical magnetic field strength and C$^{18}$O abundance in protostellar sources, 
H$^{13}$CO$^+$ is expected to be attached to the magnetic field on a 100~au scale in B335.

In non-ideal MHD theoretical calculations and simulations incorporating ambipolar diffusion, the ambipolar drift velocity starts to increase 
on a scale of tens to hundreds of au in infalling protostellar envelopes, where the magnetic field starts to decouple from the neutral gas (e.g., Fig.~7 in \citet{Krasnopolsky02}, Fig.~3 in \citet{Mellon09}, Fig.~5 in \citet{Li11}, and Fig.~1 in \citet{Zhao18}). 
The radius, where the magnetic field starts to decouple from the neutral gas, is proportional to the magnetic diffusivity of ambipolar diffusion \citep{Krasnopolsky02, Zhao16, Zhao18}. 
In the simulations with a standard Mathis-Rumpl-Nordsieck (MRN) grain size distribution \citep{Mathis77} and a typical cosmic-ray ionization rate of 10$^{-17}$~s$^{-1}$, 
the magnetic field is well coupled with the neutral gas and is dragged to the inner few hundred au or even a smaller scale in infalling envelopes  \citep{Li11, Zhao18}. 
Then the magnetic field gradually decouples from the neutral gas in the inner regions and diffuses outward to form an ambipolar diffusion shock, which efficiently decelerates the infalling and rotational motions. \citep[e.g.,][]{Krasnopolsky02, Li11}.   
In this case, the formation of a Keplerian disk with a size larger than 10~au is suppressed because a large amount of  magnetic flux is accumulated in the inner envelope and efficiently transports away the angular momentum of the infalling material \citep{Mellon09, Li11}. 
Several simulations have shown that incorporating ambipolar diffusion with a typical diffusivity does not enable large-scale Keplerian disks to form \citep[e.g.,][]{Mellon09, Li11, Zhao16, Zhao18}.  
In these simulations, the ambipolar drift velocity is lower than 0.2~km~s$^{-1}$ at a radius of 100~au, increases to 0.3~km~s$^{-1}$ at a a radius of 70~au, and is higherer than 1~km~s$^{-1}$ at radii smaller than 30~au, while the infalling velocities of the neutrals are 0.3, 0.5, and 1.5~km~s$^{-1}$ at radii of 100, 70, and 30~au, respectively, when the central protostellar mass is 0.57~$M_\sun$ \citep[Fig.~5 in][]{Li11}. 
If the ions are attached to the magnetic field, such an ambipolar drift velocity implies that the ion velocity is lower than 30\%--40\% of the neutral velocity.

On the other hand, 
the ambipolar diffusivity is enhanced by one to two orders of magnitude when small dust grains with sizes of a few to tens of nanometers are removed from the grain size distribution \citep{Dapp12, Padovani14, Zhao16}.
As a consequence, 
the magnetic field starts to decouple from the neutrals on a larger scale of several hundred or even few thousand au. 
In the simulations, the ambipolar drift velocity can be larger than 0.5~km~s$^{-1}$ at radii of 100--300~au and increase to 1~km~s$^{-1}$ at a radius of 100~au when the protostellar mass is 0.07~$M_\sun$ \citep{Zhao18}. 
This effect enables the formation of Keplerian disks with sizes of tens of au in the simulations \citep{Masson16, Zhao16, Zhao18}.  
We note that in such a paradigm with an enhanced ambipolar diffusivity, the formation of large-scale Keplerian disks with a typical mass-to-flux ratio of a few can still be suppressed when the initial rotation$\footnotemark$ of parental cores is slow with an angular velocity lower than $(5\mbox{--}8) \times 10^{-14}$~s$^{-1}$, corresponding to a ratio of rotational to gravitational energy of 0.5--1.6\% \citep{Dapp12, Tomida15, Zhao18}.

\footnotetext{The initial core rotation is assumed to be rigid-body rotation in the simulations.}

With our observations, we set an upper limit to the velocity difference between the ionized and neutral gas at a radius of 100~au to be 0.3~km~s$^{-1}$ in B335.
Our results are consistent with the non-ideal MHD simulations with the ambipolar diffusivity computed with the standard MRN grain size distribution, 
where the ambipolar drift velocity is expected to be $<$0.2~km~s$^{-1}$ at a radius of 100~au. 
Since our observations are not able to resolve the inner region of a few tens of au, 
we are not able to further unambiguously detect any increase in drift velocity toward the inner regions, which is predicted in the simulations.
On the contrary, 
our observational results of B335 are inconsistent with the non-ideal MHD simulations with the enhanced ambipolar diffusivity and the removal of the small dust grains$\footnotemark$, where the minimum grain size is 0.1 $\mu$m. 
The ambipolar drift velocity of 0.5--1~km~s$^{-1}$ on a scale of 100~au predicted in these simulations with the enhanced ambipolar diffusivity is not detected in B335 with our observations.

\footnotetext{In non-ideal MHD simulations incorporating ambipolar diffusion, the ambipolar drift velocity is expected to lower when the minimum size in the grain size distribution is smaller \citep{Zhao18}.}

Therefore, 
based on the fact that H$^{13}$CO$^+$ is likely well coupled to the magnetic field and that there is no measurable difference in the infall velocities between H$^{13}$CO$^+$ and C$^{18}$O on a 100~au scale probed by our observations, 
our results suggest that in B335 any significant magnetic decoupling from the bulk neutral matter, if present, likely occurs on a scale smaller than 100~au, as in the non-ideal MHD simulations with the typical ambipolar diffusivity.
Therefore, the magnetic field is dragged and accumulated on the small scale, and magnetic braking can efficiently remove the angular momentum from the material infalling to the small scale and suppress the disk formation in B335. 
This can result in the absence of a Keplerian disk with a size larger than 10 au in B335, as found in the observations \citep{Yen15b}. 

There are also other non-ideal MHD simulations that only incorporate Ohmic dissipation but no ambipolar diffusion \citep[e.g.,][]{Dapp10, Tomida13}. 
These simulations, which resolve the formation of first cores, show that small Keplerian disks with a size of few au can form during the phase of first cores or during the formation of second cores.  
These simulations were unable to follow the evolution of the Keplerian disks further. 
Although it is not clear whether the size of these small disks will grow to a scale of 100~au as expected \citep{Tomida15} or remains on a scale of 10~au \citep{Dapp12}, 
in these simulations the magnetic field is decoupled from the neutral gas in the innermost regions. 
B335 can also be an observational analog to these simulations. 
In addition, the non-detection of the velocity difference between the ionized and neutral gas is also consistent with the ideal MHD case. 
Nevertheless, no Keplerian disk is expected to form in the ideal MHD case unless the magnetic field is weak and largely misaligned from the rotational axis of parental cores \citep{Mellon08, Joos12, Li13}, 
while the presence of the outflow and jets in B335 suggests that there is at least a small rotating disk around the protostar \citep[e.g.,][]{Blandford82, Pudritz83, Pudritz86}. 
Thus, B335 is less likely the ideal MHD case. 
Future observations with higher angular resolutions to measure the velocity difference between the ionized and neutral gas on a scale of tens of au in B335 are needed to  distinguish these simulations and the ones with the typical ambipolar diffusivity.

We also note that the velocity drift between ionized and neutral gas is expected to be most significant in the midplane of infalling protostellar envelopes, where the magnetic field is highly pinched, and there is almost no velocity drift in the upper layers of envelopes, as discussed in \citet{Krasnopolsky02} and \citet{Zhao18}. 
Although the inclination of B335 is edge on, the most suitable case to probe the gas motions in the midplane, 
our observations may not be able to distinguish the gas motions in the midplane and upper layers in the protostellar envelope in B335 because of the limited angular resolution, 
and the signature of the velocity drift (if present) is possibly diluted and smoothed out. 
Further observations with an angular resolution of $\sim$0\farcs1 to resolve the vertical velocity structures within 10--20~au from the midplane at radii of 100--200~au are required to examine this possibility.

Finally, if the magnetic field strength on a 100~au scale in B335 is only a few tens of $\mu$G rather than $>$100~$\mu$G, as expected from the current theories and observations, 
specific ion species such as H$^{13}$CO$^+$ are not the ideal indicator of the kinematics of the magnetic field because they are detached from the magnetic field ($\beta_{i,{\rm H_2}} < 1$). 
In this case,
ions and neutrals in the inner 100~au region can move together, but the magnetic field can still be left behind the bulk infall motion. 
This picture of ions detaching from the magnetic field is also consistent with our non-detection or upper limit of the velocity difference between ions and neutrals at a radius of 100~au. 
In this case, the magnetic field is not dynamically important in the infalling protostellar envelope \citep{Mellon09}.
The Keplerian disk is expected to grow in size with the proceeding collapse \citep{Terebey84, Basu98}, 
and the absence of a Keplerian disk larger than 10~au in B335 might simply be due to its young age \citep{Yen15b}.  

\section{Summary}
We analyzed the data of the H$^{13}$CO$^+$ (3--2) and C$^{18}$O (2--1) emission in the Class 0 protostar B335 obtained with our ALMA observations. 
The goal was to investigate ambipolar diffusion in protostellar envelopes by comparing the gas motions traced by the ionized and neutral molecular lines.
Our main results are summarized below. 
\begin{enumerate}
\item{The H$^{13}$CO$^+$ and C$^{18}$O emission lines both show a central compact component with a size of 1$\arcsec$--2$\arcsec$ (100--200~au) elongated perpendicular to the outflow direction, and there are additional blueshifted extensions toward the east. Clear velocity gradients along the outflow direction are observed in both lines. The extensions toward the east are likely associated with the outflow. The elongated central components in the H$^{13}$CO$^+$ and C$^{18}$O emission likely trace the flattened envelope around the protostar, and their velocity structures can be explained with an edge-on infalling envelope.}
\item{We constructed kinematical models of an infalling and rotating envelope and fitted the observed velocity structures in the PV diagrams along the major and minor axes of the flattened envelope in the H$^{13}$CO$^+$ and C$^{18}$O emission. The infalling velocities traced by the H$^{13}$CO$^+$ and C$^{18}$O emission are both measured to be 0.85$\pm$0.2~km~s$^{-1}$ at a radius of 100~au, suggesting that the velocity difference between the ionized and neutral gas is at most 0.3~km~s$^{-1}$ at a radius of 100~au.} 
\item{The Hall parameter of H$^{13}$CO$^+$ is estimated to be $\gg$1 on a 100~au scale in B335 on the assumption of a typical magnetic field strength of $>$100~$\mu$G on a 100~au scale in protostellar envelopes. Thus, H$^{13}$CO$^+$ is expected to be attached to the magnetic field, and the difference in the infalling velocities measured in the H$^{13}$CO$^+$ and C$^{18}$O lines can trace the ambipolar drift velocity.}
\item{Our non-detection or upper limit of the ambipolar drift velocity suggests that the magnetic field and the bulk neutral matter remain well coupled on a 100~au scale, and any significant decoupling, if present, likely occurs on smaller scales than probed by our observations. Consequently, the magnetic field is dragged and accumulated on the small scale, and efficient magnetic braking can suppress the disk formation in B335, as expected in non-ideal MHD simulations with a standard MRN grain size distribution and a typical cosmic-ray ionization rate.  
On the other hand, the high ambipolar drift velocity of 0.5--1~km~s$^{-1}$ predicted in the non-ideal MHD simulations with the enhanced ambipolar diffusivity by removing the small dust grains, where the minimum grain size is 0.1 $\mu$m, is not detected in our observations.
However, we cannot rule out that there is a significant ambipolar drift only in the midplane of the infalling envelope. Such signatures could be smoothed out in our observations due to the limited angular resolution.
While the non-detection of a velocity difference is still also consistent with ideal MHD simulations, 
the existence of a small disk in B335 makes this scenario less probable.
Future observations with higher angular resolutions are needed to establish the definite presence of ambipolar diffusion in B335.}
\item{Although it is less likely, if the magnetic field strength is only tens of $\mu$G on a 100~au scale in B335, H$^{13}$CO$^+$ can be detached from the magnetic field and is no longer an indicator of the kinematics of the magnetic field. Thus, H$^{13}$CO$^+$ and the neutrals in the inner 100~au region can move together, and there would be no difference in the infalling velocities traced by the H$^{13}$CO$^+$ and C$^{18}$O emission. In this case, the magnetic field is not dynamically important in the infalling protostellar envelope.}
\end{enumerate}

\begin{acknowledgements} 
This paper makes use of the following ALMA data: ADS/JAO.ALMA\#2013.1.00879.S, ADS/JAO.ALMA\#2015.1.01188.S. ALMA is a partnership of ESO (representing its member states), NSF (USA) and NINS (Japan), together with NRC (Canada), MOST and ASIAA (Taiwan), and KASI (Republic of Korea), in cooperation with the Republic of Chile. The Joint ALMA Observatory is operated by ESO, AUI/NRAO and NAOJ. We thank all the ALMA staff supporting this work. ZYL is supported in part by NASA NNX14AB38G and NSF AST-1313083 and 1716259. PMK acknowledges support from the Ministry of Science and Technology through grant
103-2119-M-001-009 and from an Academia Sinica Career Development Award. 
ST is supported by NAOJ ALMA Scientific Research Grant Numbers 2017-04A.
ST acknowledges a grant from the Ministry of Science and Technology (MOST) of Taiwan (MOST 102-2119-M-001-012-MY3), and JSPS KAKENHI Grant Number JP16H07086, in support of this work.
\end{acknowledgements}

\begin{appendix}
\section{Position--velocity diagrams of kinematical models}\label{sigmod}
\begin{figure*}
\centering
\includegraphics[width=18.5cm]{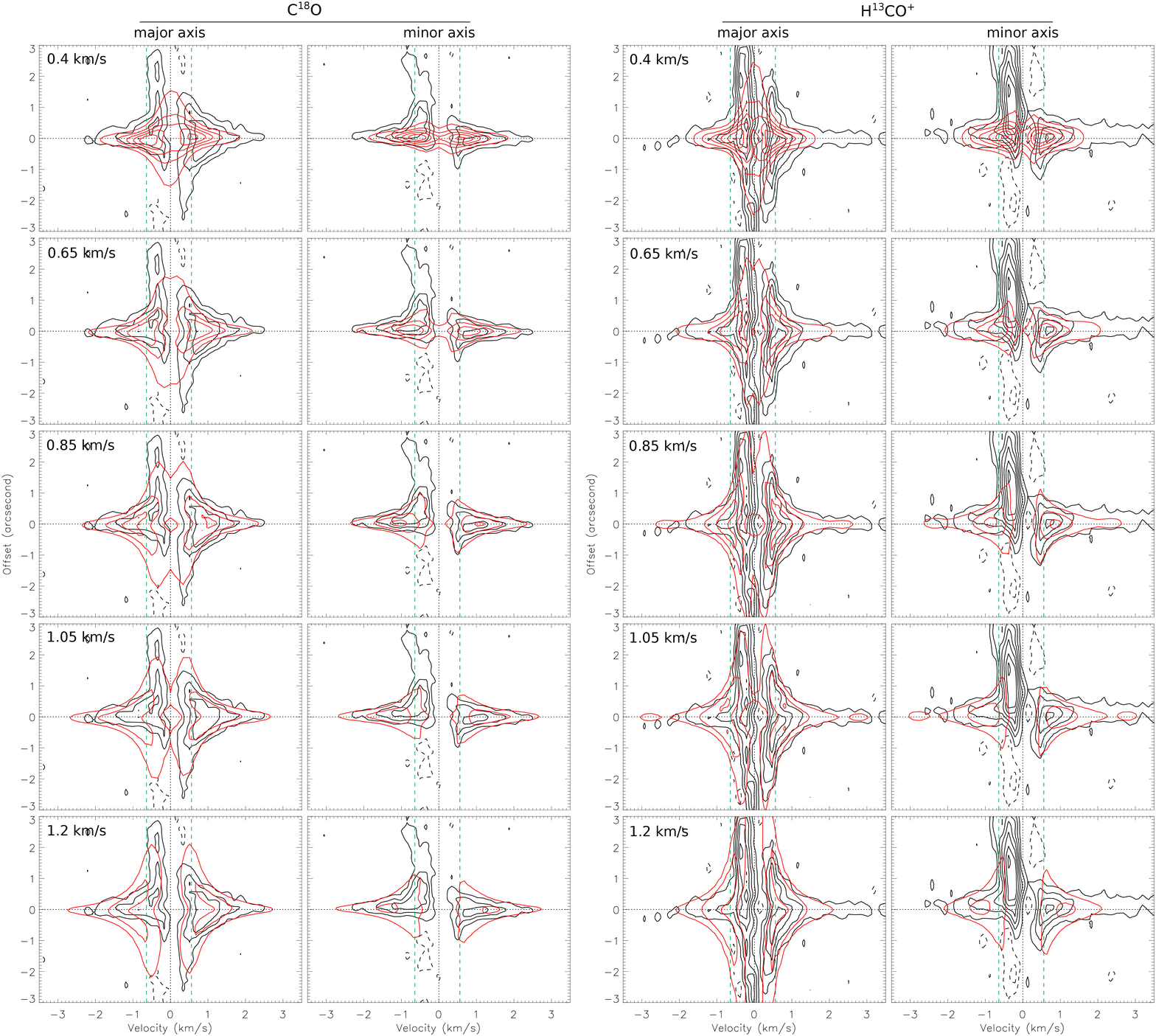}
\caption{Same as Fig.~\ref{modelpv}, but for comparison with models with different $V_{\rm in}$(100~au) for C$^{18}$O (left columns) and H$^{13}$CO$^+$ (right columns),  and $V_{\rm in}$(100~au) of each model is labeled at the upper left corner in the panel.}\label{sigpv}
\end{figure*}

Figure \ref{sigpv} compares the observed PV diagrams with models with different $V_{\rm in}$(100~au) from 0.4~km~s$^{-1}$ to 1.2~km~s$^{-1}$. 
For these models, all the other parameters, $n(R_0)$ and $p$, were varied to minimize $\chi^2$ with the given $V_{\rm in}$(100~au).
This comparison shows that when $V_{\rm in}$(100~au) is beyond the range of 0.85$\pm$0.2~km~s$^{-1}$, the line widths in the models become wider or narrower than the observations for both C$^{18}$O and H$^{13}$CO$^+$, especially at offsets from 0\farcs5 to 1\arcsec, and the velocities of the intensity peaks in the model PV diagrams are offset from the observations. 
These results suggest that the infalling velocities of the C$^{18}$O and H$^{13}$CO$^+$ gas at a radius of 100~au are most likely within the range of 0.85$\pm$0.2~km~s$^{-1}$.

\end{appendix}

\end{document}